\documentclass[reprint,superscriptaddress,showpacs,amsmath,amssymb,aps,prl]{revtex4-1}
\usepackage{graphicx}
\usepackage{dcolumn}
\usepackage{bm}
\usepackage{hyperref}
\usepackage{subfigure}
\hypersetup{colorlinks=true, citecolor=blue, urlcolor=blue, linkcolor=blue}
\begin{document}
\title{Machine learning phases of active matter}

\author{Tingting Xue}\thanks{Equal contribution}
\affiliation{School of Systems Science, Beijing Normal University,
Beijing 100878, P. R. China}
\author{Xu Li}\thanks{Equal contribution}
\affiliation{School of Systems Science, Beijing Normal University,
Beijing 100878, P. R. China}
\author{Xiaosong Chen}
\affiliation{School of Systems Science, Beijing Normal University,
Beijing 100878, P. R. China}
\author{Li Chen}
\email[Email address: ]{chenl@snnu.edu.cn}
\affiliation{School of Physics and Information Technology, Shaanxi Normal University,
Xi'an 710061, P. R. China}
\author{Zhangang Han}
\email[Email address: ]{zhan@bnu.edu.cn}
\affiliation{School of Systems Science, Beijing Normal University,
Beijing 100878, P. R. China}

\begin{abstract}
Recent years have witnessed a growing interest in using machine learning to predict and identify phase transitions in various systems. Here we adopt convolutional neural networks (CNNs) to study the phase transitions of Vicsek model, solving the problem that traditional order parameters are insufficiently able to do.  Within the large-scale simulations, there are four phases, and we confirm that all the phase transitions between two neighboring phases are first-order.
We have successfully classified the phase by using CNNs with a high accuracy and identified the phase transition points, while traditional approaches using various order parameters fail to obtain.
These results indicate that the great potential of machine learning approach in understanding the complexities in collective behaviors, and in related complex systems in general.
\end{abstract}

\date{\today }
\maketitle

\emph{Introduction} --- Active matter emerged as a novel class of non-equilibrium systems, in which individual components absorb energy from their surroundings and, in some way, transform it into mechanical work~\cite{netzer2019heterogeneous, marchetti2013hydrodynamics, ramaswamy2017active}.
Examples are ubiquitous in nature, ranging from bacteria colonies, cells~\cite{wu2000particle, elgeti2015physics, prost2015active}, bird flocks, to fish schools~\cite{cavagna2014bird, calovi2014swarming}, and human crowds~\cite{bottinelli2016emergent, bain2019dynamic}.
These systems have been the subject of numerous experimental and theoretical investigations in recent years.
Due to the grand challenge in understanding the non-equilibrium nature of active matter, several minimal models have been proposed in recent years to capture the physical principles therein~\cite{shaebani2020computational}.
Among them, the Vicsek model is the most well-known~\cite{vicsek1995novel}, and has considerably advanced our understanding in the physics behind. It models the swarming population as a group of self-propelled particles, the qualitative change in the collective behaviors is described by the phase transitions (PTs). The nature of order-disorder PT, however, was once controversial. The early studies claimed that this PT is similar to an equilibrium continuous transition, but later this conclusion was challenged by large-scale simulations and is confirmed that it is discontinuous, like a liquid-gas transition with phase separation~\cite{solon2015phase, solon2015pattern}.
In fact, there are three rather than simply two phases (ordered and disordered), i.e. the disordered gas-like state, the ordered liquid-like state, and the coexistence of ordered bands and disordered gas due to the density-velocity feedback~\cite{chate2008collective, xue2020swarming}.
More recently, with Eigen microstates approach, Ref.~\cite{li2021discontinuous} revealed that the phase transition in the Vicsek model is essentially a hybrid of continuous and discontinuous PTs, with the discontinuous component dominating as the system size goes infinity. This new perspective well reconciles and terminates the debates of the order-disorder PT.

Interestingly, the latest study~\cite{kursten2020dry} identified a fourth phase, i.e. the cross sea phase, in the ordered region, which is self-organized, as there is no external driving force. This new phase is not simply a superposition of two band waves, but an independent complex pattern with an inherently selected crossing angle. Therefore, there are four different phases identified so far in the Vicsek model: the ordered phase, the cross sea phase, the band phase, and the disordered phase. Precisely identifying these phase transition points, however, now becomes extremely challenging.

\begin{figure*}[t]
\centering
{\includegraphics[width=15cm]{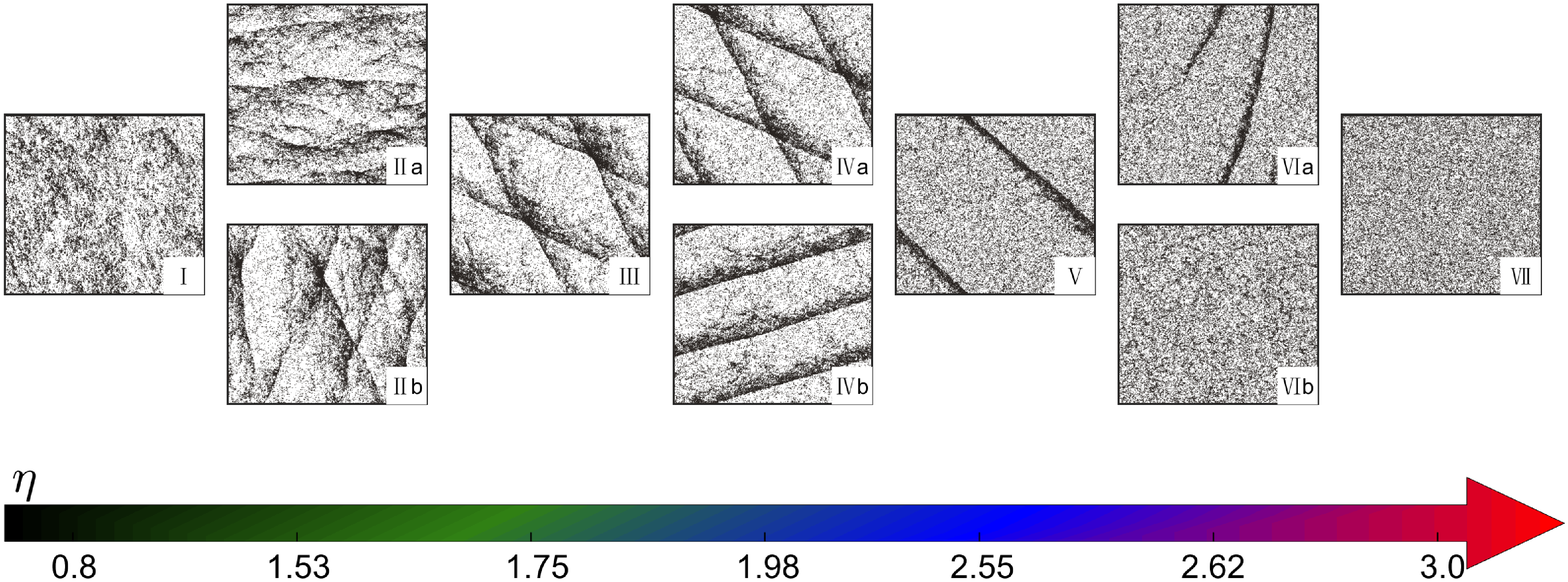}}
\caption{(Color online) Typical snapshots for the Vicsek model at different noise amplitudes.
{\small$\uppercase\expandafter{\romannumeral1}$}: the ordered phase at $\eta=0.8$;
{\small$\uppercase\expandafter{\romannumeral2}$}: a bistable state at $\eta=1.53$ ((a) the ordered phase, and (b) the cross sea phase);
{\small$\uppercase\expandafter{\romannumeral3}$}: the purely cross sea phase at $\eta=1.75$;
{\small$\uppercase\expandafter{\romannumeral4}$}: a bistable state at $\eta=1.98$ ((a) the cross sea phase, and (b) the band phase);
{\small$\uppercase\expandafter{\romannumeral5}$}: purely band phase at $\eta=2.55$;
{\small$\uppercase\expandafter{\romannumeral6}$}: a bistable state at $\eta=2.62$ ((a) the band phase and (b) the disordered phase);
and {\small$\uppercase\expandafter{\romannumeral7}$}: the disordered phase at $\eta=3.0$.
Parameters: $\rho=0.5$, $v_{0}=1.0$, $L=512$.
}
\label{Fig.1}
\end{figure*}

Identifying phase transitions is to pin down boundaries between different phases of systems, a fundamental task in understanding the physics of complex systems.
Traditionally, the polar order parameter is computed to classify the different phases, which is often useful, e.g. in studying the ordered-disordered phases transition~\cite{vicsek1995novel}.
However, no jump in the average polar order parameter is seen at the transition towards the band-disordered phases, and it's also found insensitive around the cross sea phase~\cite{kursten2020dry}. Therefore, other quantities such as the Binder cumulant~\cite{chate2008collective} and the structural order parameter are proposed~\cite{kursten2020dry}, helping to better distinguish different phases.
But it is not a prior clear how to properly choose the right order parameters for many non-equilibrium systems without a trial-and-error process.
Comparatively, the identification of different phases according to typical snapshots is more accurate, but this classification approach by human eyes has its limitations when the number of snapshots is huge, but is usually required to precisely determine the regions of different phases~\cite{kursten2020dry, kursten2020multiple}.

In recent years, machine learning flourishes for its great momentum into almost every branch of science~\cite{von2020retrospective, jones2017machine, carleo2019machine}. Not surprisingly, machine learning methods have been proposed to study the matter phases~\cite{jordan2015machine, van2017learning, venderley2018machine}.
The general idea behind several works~\cite{carrasquilla2017machine, deng2017quantum,rodriguez2019identifying, wetzel2017unsupervised, zhang2019machine} is to use supervised learning; sampling the physical states for several phases, put these labeled configuration information as the training data set to train the neural networks; once the training is completed, the neural networks can be used to estimate the probabilities of testing data that belong to different phases; finally, the phase transitions can be inferred based on these probabilities.
Machine learning is not only very effective in studying phase transitions in classic models, but also has remarkable ability to reveal various phases of matter in quantum many-body systems~\cite{torlai2018latent,ch2017machine, rem2019identifying, carleo2017solving}. The current studies indicate its great potential in identifying phase transitions in complex systems, for its extraordinary advantages over traditional methods.

In this work, we employ the convolutional neural network (CNN)~\cite{krizhevsky2012imagenet,carrasquilla2017machine,carleo2019machine} to classify different phases in Vicsek model according to the sampled snapshots.
By large-scale simulations of Vicsek model, we first numerically confirm that all phase transitions in each of the two neighboring phases are first-order.
By adopting various order parameters to analyze the phase transitions, we find that the boundaries for different phases are fuzzy due to the insensitivity of these order parameters, and even the types of phase transitions cannot be determined.
But the supervised learning based on CNNs conquers these difficulties, which successfully classifies all four phases.
The precision of phase classification is robust to the choice of order parameters.

\begin{figure}[t]
\centering
{\includegraphics[width=4.2cm]{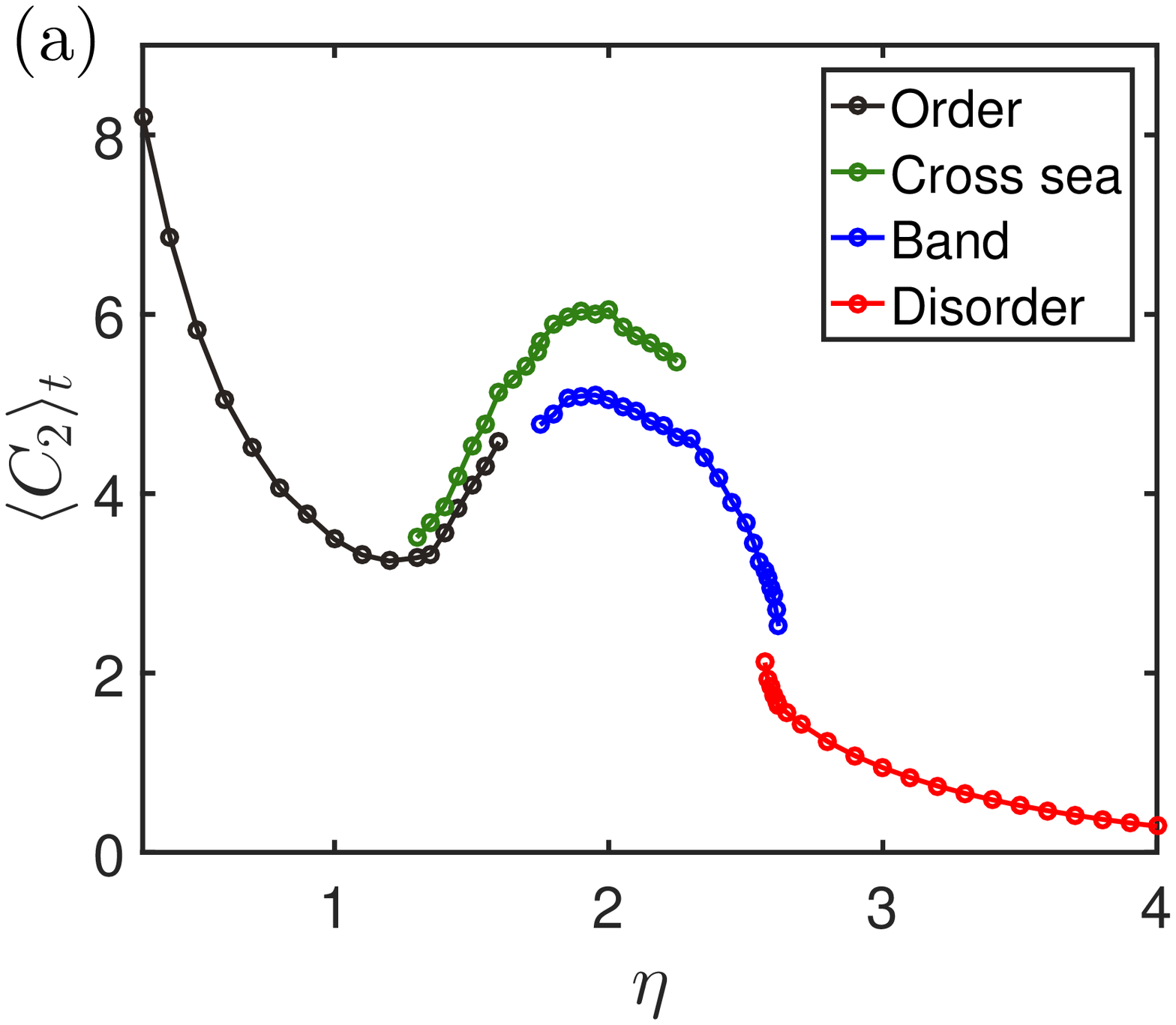}}
{\includegraphics[width=4.2cm]{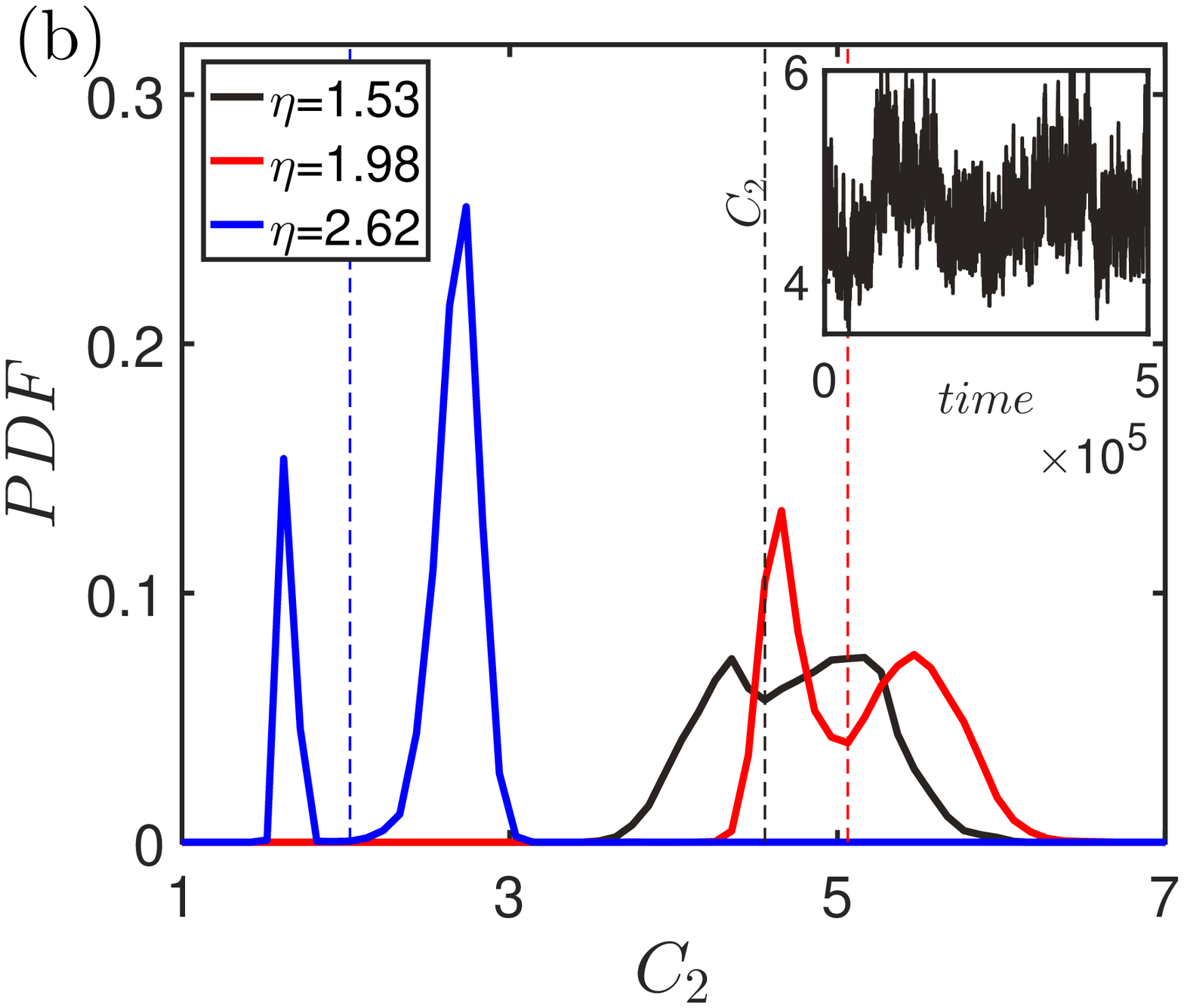}}
\caption{(Color online) Phase transitions of the Vicsek model.
(a) The structural order parameter $\langle C_{2} \rangle_t$ versus the noise amplitude $\eta$.
$\langle C_{2} \rangle_t$ is computed as follows: if the probability distribution function (PDF) of $C_{2}$ is single-peaked, then we average all data; but if the profile is bimodal, as cases shown in (b), we divide the data at the valley point, and do average respectively for the bistable states, two points are then obtained for each $\eta$. The plot shows that three transitions among the four states are all bistable, thus are of first-order nature.
(b) PDF of $C_{2}$ for the three parameters used in Fig.~\ref{Fig.1}. The double peak structures show the discontinuous nature of the transition. The inset shows the time series at the noise amplitude $\eta=1.53$.
The data is averaged over $5\times10^{5}$ time steps after the transient $5\times10^{5}$.
Parameter: $L=512$.
}
\label{Fig.2}
\end{figure}

\begin{figure*}[t]
\centering
{\includegraphics[width=18cm]{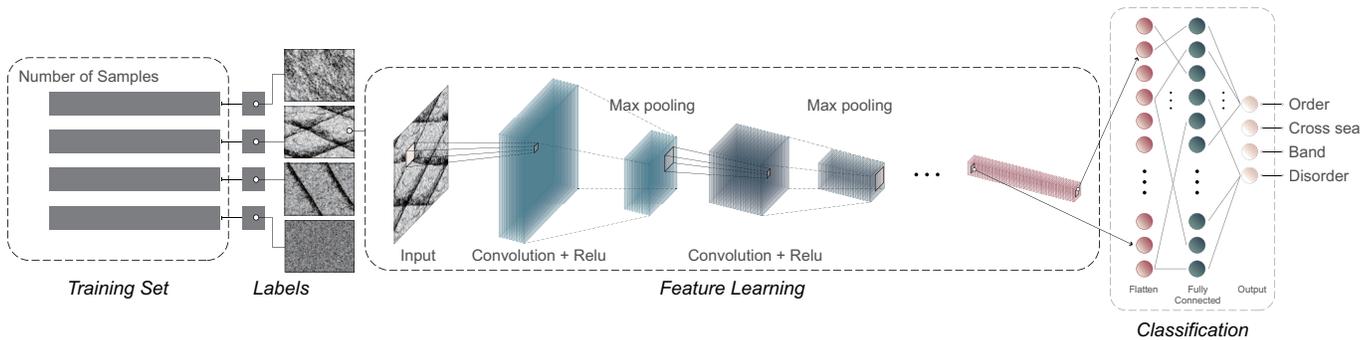}}
\caption{(Color online) A schematic diagram of the convolutional neural network for the phase classification.
}
\label{Fig.3}
\end{figure*}

\emph{The model} --- In the standard Vicsek model~\cite{vicsek1995novel}, $N$ self-propelled particles with a constant velocity $v_{0}$ that move in a two-dimensional domain of size $L\times L$ with the periodic boundary condition. They move synchronously at discrete time steps by a fixed distance $v_{0}\Delta t$. Each particle $i$ is endowed with an angle $\theta_{i}$ that determines the movement direction. Its update is determined by the average orientations of its neighbors (defined as particles within a circle of radius $R=1$ centered around particle $i$, including itself) plus some noise. Specifically,
\begin{equation}\large
	\theta_{i}(t+\Delta t)=\Theta[\sum_{j:\;d_{ij}<R} \mathbf{v}_j(t)]+\eta \xi _{i}(t) ,
\end{equation}
where $\Theta[\mathbf{v}]$ represents the angle that describes the direction of the two dimensional vector $\mathbf{v}$ and $d_{ij}$ is the distance between particle $i$ and $j$. The noise $\xi _{i}$ is chosen from the uniform distribution within the interval $[-1/2, 1/2]$, and $\eta$ is the noise amplitude. In our study, the density $\rho=0.5$ and $v_0=1.0$ are fixed throughout the study.

\emph{Results and analysis} ---
We first report typical snapshots in Fig.~\ref{Fig.1} to provide the overall picture for the Vicsek model when the noise is varied.
For small noise, the ordered phase is seen in Fig.~\ref{Fig.1}{\small($\uppercase\expandafter{\romannumeral1}$)}. Increasing noise leads to the cross sea phase [Fig.~\ref{Fig.1}{\small($\uppercase\expandafter{\romannumeral3}$)}], where the interactions become more intensive for the cluster characteristics. With further increase in noise amplitude $\eta$, the band structure could emerge [Fig.~\ref{Fig.1}{\small($\uppercase\expandafter{\romannumeral5}$)}], which is locally ordered and of high density. Finally, the system evolves into the disordered phase when the noise become very strong, see Fig.~\ref{Fig.1}{\small($\uppercase\expandafter{\romannumeral7}$)}.
However, we find that the phase boundaries are not distinct, bistable states are always seen for each two neighboring phases.
For example, snapshots in Fig.~\ref{Fig.1}{\small($\uppercase\expandafter{\romannumeral2}$)} show the coexistence of ordered and cross sea phases, where the two phases interchange into each other from time to time. Similar observations are also made between the cross sea and the band phase [Fig.~\ref{Fig.1}{\small($\uppercase\expandafter{\romannumeral4}$)}], and between the band phase and the disordered phase [Fig.~\ref{Fig.1}{\small($\uppercase\expandafter{\romannumeral6}$)}]. Here, the bistablity and the phase coexistence in Fig.~\ref{Fig.1}{\small($\uppercase\expandafter{\romannumeral6}$)} are confirmed in previous studies~\cite{chate2008collective}, which is a first-order phase transition; but the nature of other two transitions [from {\small$\uppercase\expandafter{\romannumeral1}$} to {\small$\uppercase\expandafter{\romannumeral3}$}, {\small$\uppercase\expandafter{\romannumeral3}$} to {\small$\uppercase\expandafter{\romannumeral5}$}], to our best knowledge, have not yet been proved.

Because of the insensitivity of the polar order parameter to the phase transition, especially around the cross sea and band phases, here we adopt the structure order parameter $C_{2}$~\cite{kursten2020multiple, kursten2020dry} as our new order parameter. It is a local integral over the two particle correlation function formally given by
\begin{equation}\large
\begin{split}
	C_{2}=(\frac{N}{L^{2}})^{2} \int _{R^{2}}[g(|\mathbf{r_{1}}-\mathbf{r_{2}}|)-1]\\
 \times \theta (R-|\mathbf{r_{1}}|) \theta (R-|\mathbf{r_{2}}|)d \mathbf{r_{1}}d\mathbf{r_{2}},
\end{split}
\end{equation}
where $\theta$ is the Heaviside function, $g(r)$ is the usual pair correlation function. The value is zero for the independent particle ensemble, but becomes large when particles are clustered. Therefore, $C_2$ works generally better than the traditional polar order parameter and the Binder cumulant for its ability in capturing the structural change. A simple way of computing $C_2$ is given in Sec. \uppercase\expandafter{\romannumeral1} of SM~\cite{SM}.

Fig.~\ref{Fig.2}(a) shows the dependence of the average structure order parameter $C_{2}$ on the noise amplitude, where two cases can be seen. While a single value of $C_{2}$ corresponds to a pure state [e.g. Fig.~\ref{Fig.1}{\small($\uppercase\expandafter{\romannumeral1}, \uppercase\expandafter{\romannumeral3},\uppercase\expandafter{\romannumeral5},\uppercase\expandafter{\romannumeral7}$)}], the two-value cases are for the bistable states [Fig.~\ref{Fig.1}{\small($\uppercase\expandafter{\romannumeral2}, \uppercase\expandafter{\romannumeral4},\uppercase\expandafter{\romannumeral6}$)}]. We can see that the bistablity is observed for all three transitions among four pure states, indicating first-order phase transitions. And this observation is strengthened by the bimodal PDF as well as the time series, see Fig.~\ref{Fig.2}(b)

We also compare the structure order parameter $C_2$ with the traditional polar order parameter, and find that $C_2$ indeed has advantage in distinguishing the cross sea phase (for details see Sec. \uppercase\expandafter{\romannumeral2} of SM~\cite{SM}). Still, the change in $C_{2}$ is very small in some certain bistable states, or the system is prone to stabilize into a state within the bistable region with a little chance to evolve into the other due to the deep potential well.
All these factors lead to great difficulties in the phase segmentation. In addition, both order parameters are greatly influenced by the system size (see Sec. \uppercase\expandafter{\romannumeral3} of SM~\cite{SM}).

\begin{figure*}
\centering
{\includegraphics[width=6.0cm]{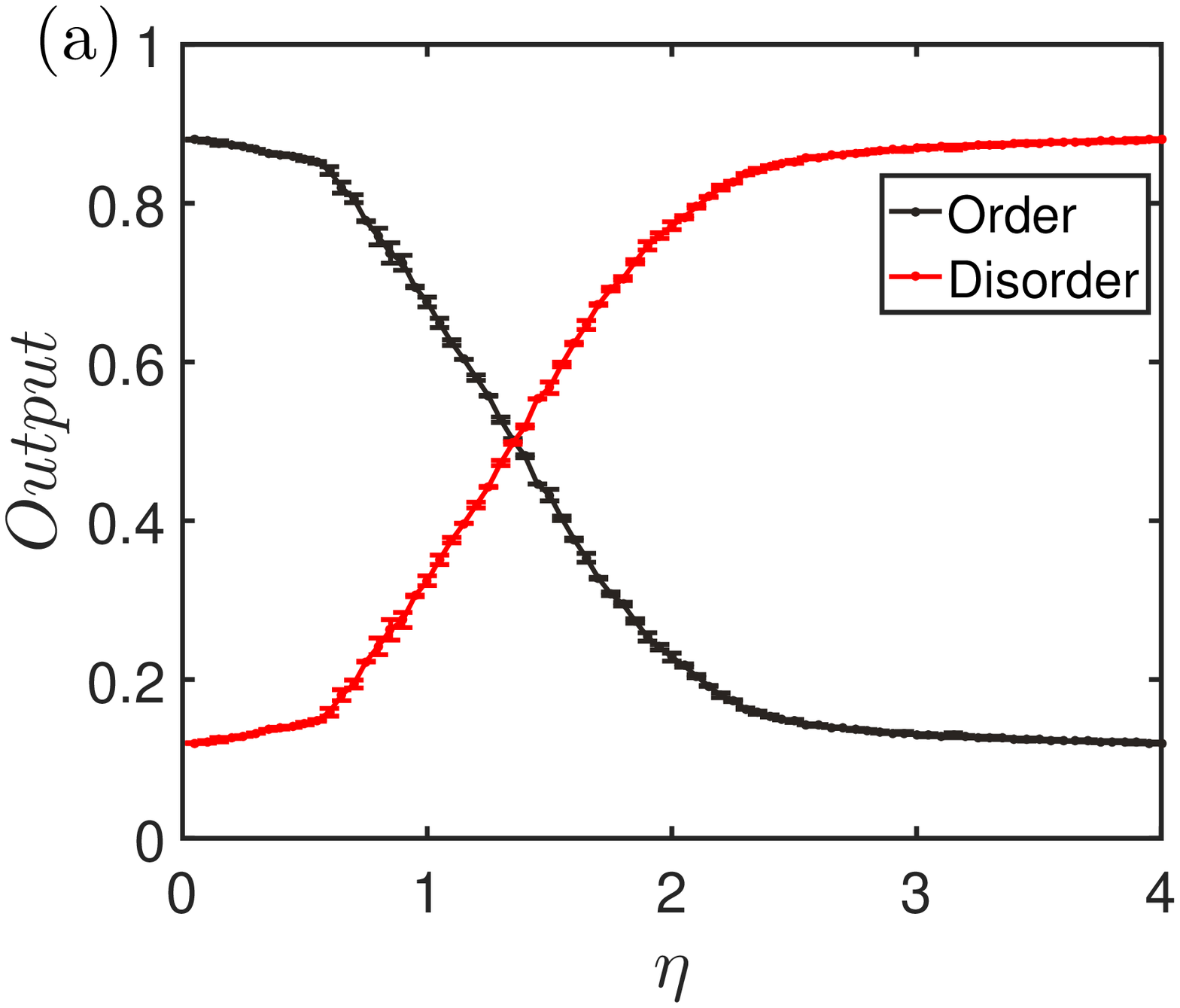}}
{\includegraphics[width=6.0cm]{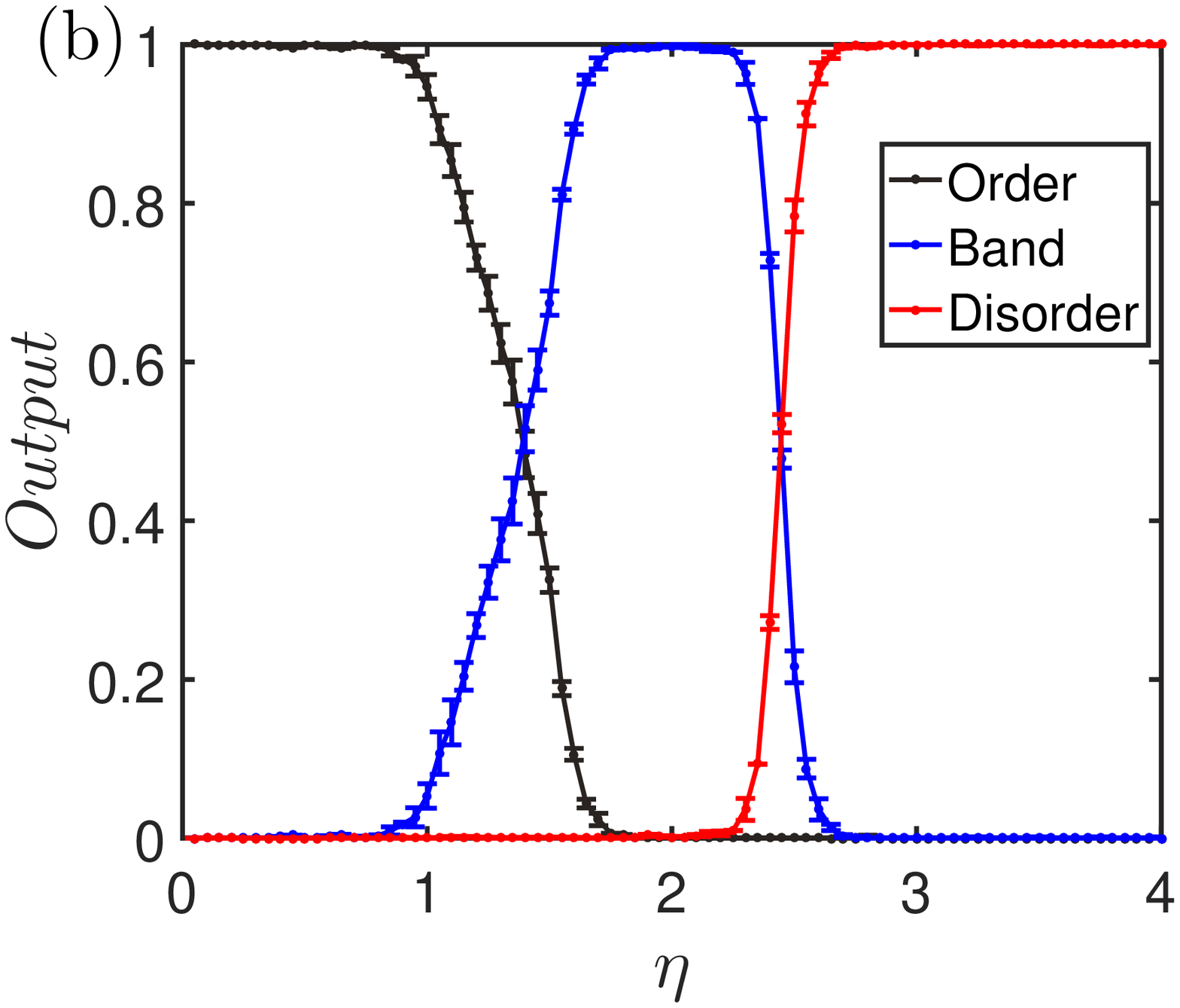}}
{\includegraphics[width=6.0cm]{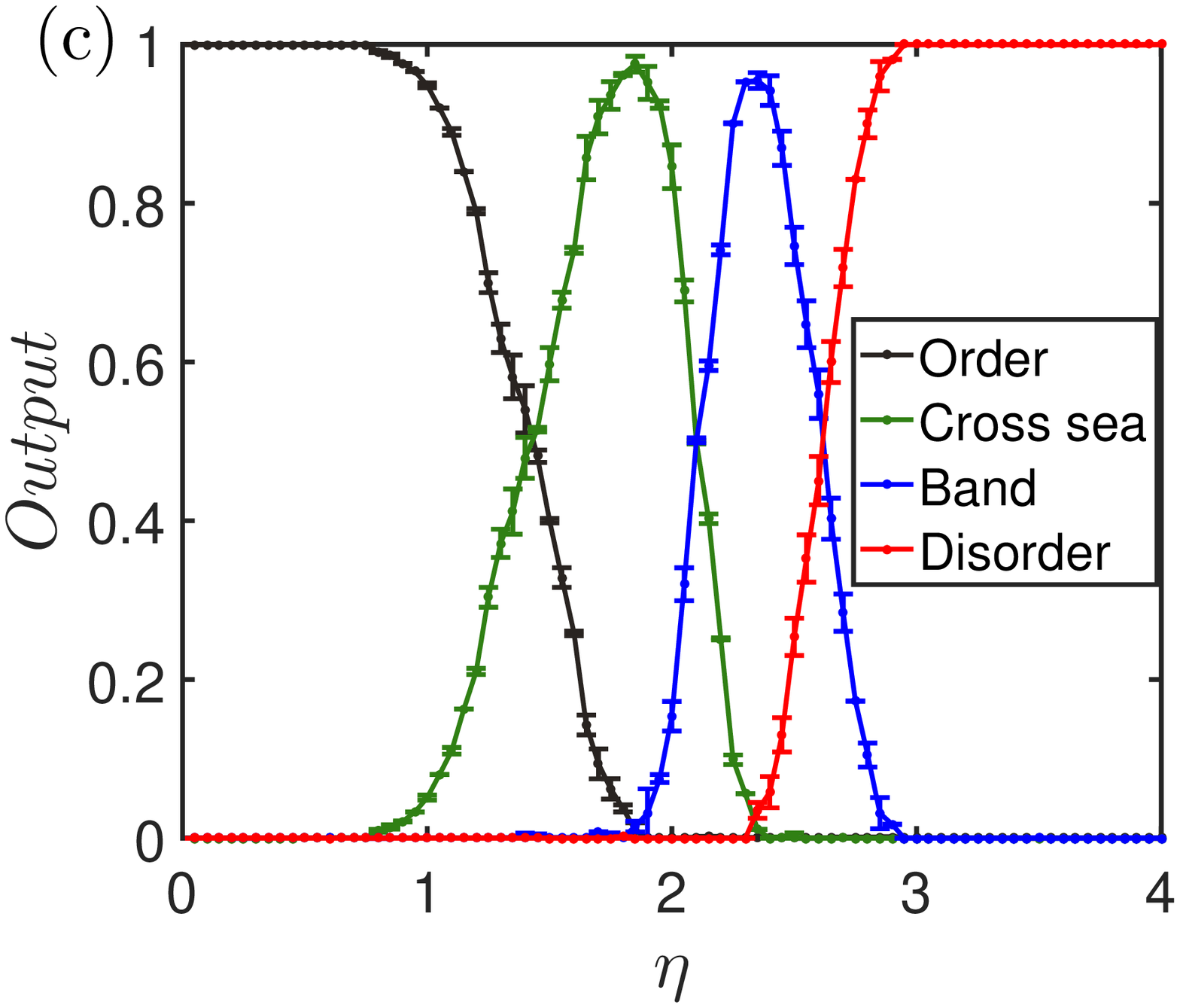}}
{\includegraphics[width=6.0cm]{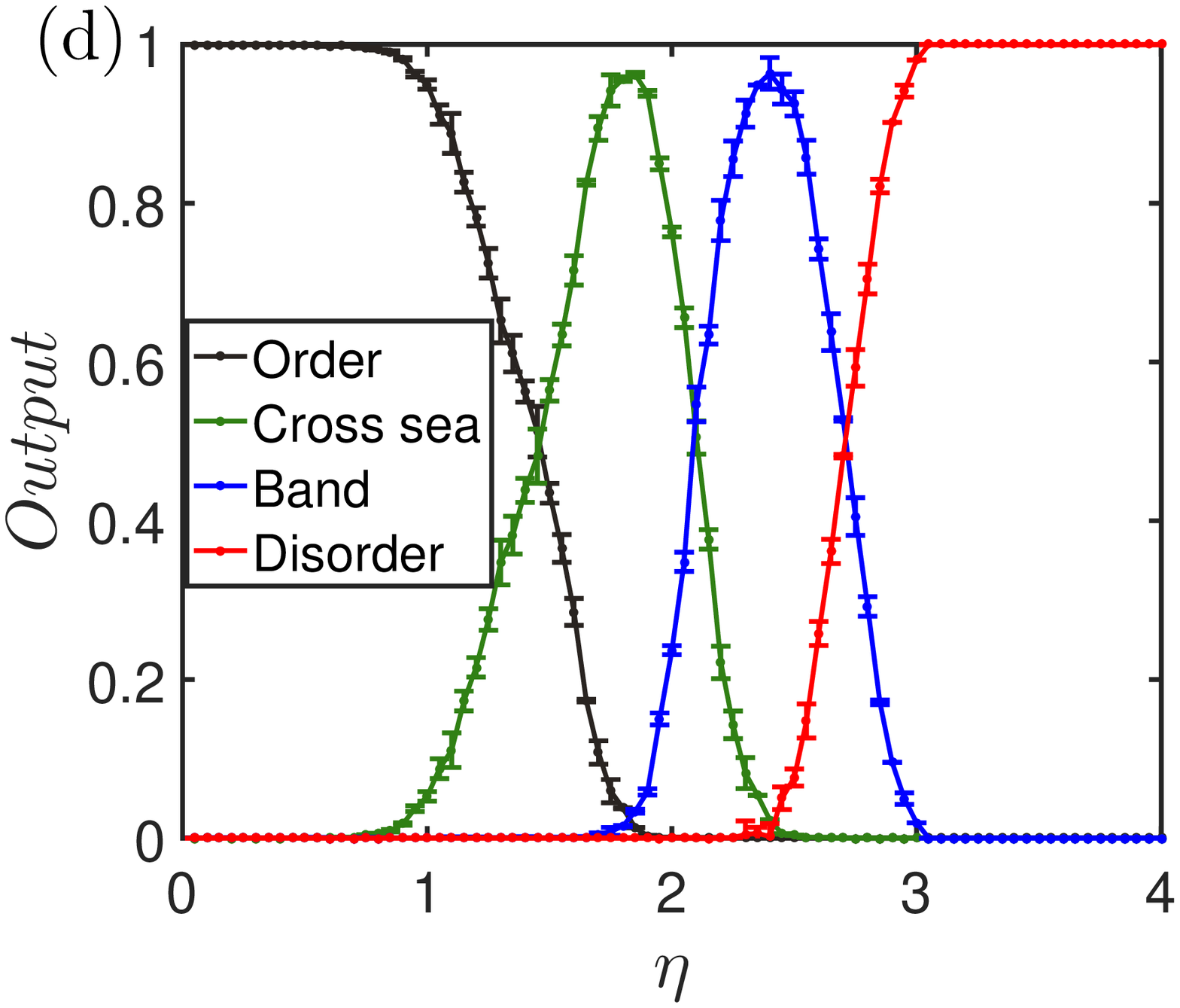}}
\caption{(Color online) Machine learning different phases in the Vicsek model.
The output on the test set versus the noise amplitude $\eta$ for the system with size $L=32, 128, 512$ and 1408, respectively for (a) to (d). Only two states are presented in the small system (a), and the emergence of the cross sea phase requires the system is large enough (c,d).
The output layer is averaged $10$ realizations, the error bars represent the standard deviation.
}
\label{Fig.4}
\end{figure*}

\emph{Convolutional neural network} ---
We now turn to adopt CNNs to classify the rich phases of Vicsek model with the help of TensorFlow~\cite{abadi2015tensorflow}.
The particle configuration is sampled as an image, and the task turns to be the image classification.
The method belongs to the supervised learning that try to find the correct mapping between the configuration and the phase label.
Fig.~\ref{Fig.3} shows the network structure of the CNN. It is composed of an input layer with the same size as the image pixel sampled, multiple cascaded convolution-pooling layers, fully connected layers, and the output layer~\cite{lecun2015deep, goodfellow2016deep}.
Each convolution-pooling stage is a series of the convolution layer to extract a feature map and the max-pooling layer to reduce the feature map size.
We use zero padding to maintain the spatial dimensions of feature maps during the convolution processes~\cite{krizhevsky2012imagenet, goodfellow2016deep}.
To avoid a vanishing gradient problem during training, we use the rectified linear unit (ReLU) activation for each layer of convolutional neural network.
We utilize the stochastic gradient descent optimization function in the learning rate for stable convergence~\cite{shalev2013stochastic}.

In our experiment, we use three independent data sets: the training, validation, and test sets. The training and validation sets are sampled far from the three bistable regions to guarantee the correctness for the assigned label for each image, while the test set is prepared across the whole parameter domain without label (more details see Sec.\uppercase\expandafter{\romannumeral3} of SM~\cite{SM}).
For details about CNNs and the data sets, see Sec.\uppercase\expandafter{\romannumeral4} of SM~\cite{SM}.

The whole training process consists of two stages. In the first stage, in each epoch the CNN is fed with all training data, and then we validate the CNN with the validation set to monitor its performance and stability.
By validation, we find the CNNs are able to classify different phases successfully, the accuracy is almost $100 \%$ when the system size is not very large. This is beyond our expectation because when the system is small, the snapshots are difficult to distinguish artificially by eyes.
When the system becomes large, the CNN still can classify the ordered, cross sea, band and disordered phases. Though the accuracy now decreases to around $86.2\%$. This is due to the narrow regions for the newly emerging phases, the correctness of labels in the training data sets cannot be guaranteed, thus leading to the accuracy decrease on the validation set. More details see Sec. \uppercase\expandafter{\romannumeral4} of SM~\cite{SM}.

After the training is completed, we enter into the second stage, use the trained CNNs to classify the images in the test data set.
The output probabilities for the available phases on the test sets as a function of the noise $\eta$ are shown in Fig.~\ref{Fig.4}, for four different system sizes.
Given these probabilities, their intersections can naturally interpreted as the transition point of two neighboring phases, since the probabilities there are approximately $50\%$. With this idea, we estimate $\eta_{c_{1}}\approx 1.47$, $\eta_{c_{2}}\approx 2.11$, $\eta_{c_{3}}\approx 2.58$ in Fig.~\ref{Fig.4} (c), all fall within the bistable regions. It's worthy to note that in the first-order phase transitions, the boundary for two neighboring phases is a parameter interval defined by two thresholds.  Our estimated transition point instead provides a single threshold that approximately divide the two phase regions.

\emph{Conclusions }--- Alongside the previous concerns, here we aim to clarify the nature of all phase transitions in the Vicsek model. By adopting the structural order parameter~\cite{kursten2020dry} and the technique of PDF segmentation~\cite{xue2020swarming}, we confirm that all phase transitions between the two neighboring phase are first-order given the system is large enough. The existing problem is that the structural order parameter is still insensitive to the parameter change in some regions, that makes the PDF segmentation bad and thus less precise in characterizing the phase transitions.
Therefore, we use the CNNs to classify different phases by feeding them labeled snapshot images. We find that this approach is able to successfully do phase classification with high accuracies. The transition points of two neighboring phases can also be inferred accordingly.

Our approach is feature extraction of raw data from different states, which not only correctly classifies the given states into different phases, but also infers phase transition points as well. This work opens up new possibilities for exploring phase transitions in systems where a priori knowledge is unavailable, thus the proper order parameter is unknown.
Given the success~\cite{ni2019machine, rem2019identifying, li2021determining} in complex networks, materials science, quantum systems etc, we look forward to applying machine learning to detect, predict and identify phase transitions in active matters and in many other non-equilibrium fields in the future.

\begin{acknowledgements}
We acknowledge Weiran Cai (Soochow University) and Jiqiang Zhang (Ningxia University) for helpful comments. Xiaosong Chen and Xu Li are supported by the National Natural Science Foundation of China under Grant no. 12135003. Zhangang Han and Tingting Xue are supported by the National Natural Science Foundation of China under Grant no. 62176022. Li Chen is supported by the National Natural Science Foundation of China under Grant no. 12075144.

\end{acknowledgements}

\bibliography{bibfile}

\begin{thebibliography}{42}%
\makeatletter
\providecommand \@ifxundefined [1]{%
 \@ifx{#1\undefined}
}%
\providecommand \@ifnum [1]{%
 \ifnum #1\expandafter \@firstoftwo
 \else \expandafter \@secondoftwo
 \fi
}%
\providecommand \@ifx [1]{%
 \ifx #1\expandafter \@firstoftwo
 \else \expandafter \@secondoftwo
 \fi
}%
\providecommand \natexlab [1]{#1}%
\providecommand \enquote  [1]{``#1''}%
\providecommand \bibnamefont  [1]{#1}%
\providecommand \bibfnamefont [1]{#1}%
\providecommand \citenamefont [1]{#1}%
\providecommand \href@noop [0]{\@secondoftwo}%
\providecommand \href [0]{\begingroup \@sanitize@url \@href}%
\providecommand \@href[1]{\@@startlink{#1}\@@href}%
\providecommand \@@href[1]{\endgroup#1\@@endlink}%
\providecommand \@sanitize@url [0]{\catcode `\\12\catcode `\$12\catcode
  `\&12\catcode `\#12\catcode `\^12\catcode `\_12\catcode `\%12\relax}%
\providecommand \@@startlink[1]{}%
\providecommand \@@endlink[0]{}%
\providecommand \url  [0]{\begingroup\@sanitize@url \@url }%
\providecommand \@url [1]{\endgroup\@href {#1}{\urlprefix }}%
\providecommand \urlprefix  [0]{URL }%
\providecommand \Eprint [0]{\href }%
\providecommand \doibase [0]{http://dx.doi.org/}%
\providecommand \selectlanguage [0]{\@gobble}%
\providecommand \bibinfo  [0]{\@secondoftwo}%
\providecommand \bibfield  [0]{\@secondoftwo}%
\providecommand \translation [1]{[#1]}%
\providecommand \BibitemOpen [0]{}%
\providecommand \bibitemStop [0]{}%
\providecommand \bibitemNoStop [0]{.\EOS\space}%
\providecommand \EOS [0]{\spacefactor3000\relax}%
\providecommand \BibitemShut  [1]{\csname bibitem#1\endcsname}%
\let\auto@bib@innerbib\@empty
\bibitem [{\citenamefont {Netzer}\ \emph {et~al.}(2019)\citenamefont {Netzer},
  \citenamefont {Yarom},\ and\ \citenamefont
  {Ariel}}]{netzer2019heterogeneous}%
  \BibitemOpen
  \bibfield  {author} {\bibinfo {author} {\bibfnamefont {G.}~\bibnamefont
  {Netzer}}, \bibinfo {author} {\bibfnamefont {Y.}~\bibnamefont {Yarom}}, \
  and\ \bibinfo {author} {\bibfnamefont {G.}~\bibnamefont {Ariel}},\
  }\href@noop {} {\bibfield  {journal} {\bibinfo  {journal} {Physica A:
  Statistical Mechanics and its Applications}\ }\textbf {\bibinfo {volume}
  {530}},\ \bibinfo {pages} {121550} (\bibinfo {year} {2019})}\BibitemShut
  {NoStop}%
\bibitem [{\citenamefont {Marchetti}\ \emph {et~al.}(2013)\citenamefont
  {Marchetti}, \citenamefont {Joanny}, \citenamefont {Ramaswamy}, \citenamefont
  {Liverpool}, \citenamefont {Prost}, \citenamefont {Rao},\ and\ \citenamefont
  {Simha}}]{marchetti2013hydrodynamics}%
  \BibitemOpen
  \bibfield  {author} {\bibinfo {author} {\bibfnamefont {M.~C.}\ \bibnamefont
  {Marchetti}}, \bibinfo {author} {\bibfnamefont {J.-F.}\ \bibnamefont
  {Joanny}}, \bibinfo {author} {\bibfnamefont {S.}~\bibnamefont {Ramaswamy}},
  \bibinfo {author} {\bibfnamefont {T.~B.}\ \bibnamefont {Liverpool}}, \bibinfo
  {author} {\bibfnamefont {J.}~\bibnamefont {Prost}}, \bibinfo {author}
  {\bibfnamefont {M.}~\bibnamefont {Rao}}, \ and\ \bibinfo {author}
  {\bibfnamefont {R.~A.}\ \bibnamefont {Simha}},\ }\href@noop {} {\bibfield
  {journal} {\bibinfo  {journal} {Reviews of modern physics}\ }\textbf
  {\bibinfo {volume} {85}},\ \bibinfo {pages} {1143} (\bibinfo {year}
  {2013})}\BibitemShut {NoStop}%
\bibitem [{\citenamefont {Ramaswamy}(2017)}]{ramaswamy2017active}%
  \BibitemOpen
  \bibfield  {author} {\bibinfo {author} {\bibfnamefont {S.}~\bibnamefont
  {Ramaswamy}},\ }\href@noop {} {\bibfield  {journal} {\bibinfo  {journal}
  {Journal of Statistical Mechanics: Theory and Experiment}\ }\textbf {\bibinfo
  {volume} {2017}},\ \bibinfo {pages} {054002} (\bibinfo {year}
  {2017})}\BibitemShut {NoStop}%
\bibitem [{\citenamefont {Wu}\ and\ \citenamefont
  {Libchaber}(2000)}]{wu2000particle}%
  \BibitemOpen
  \bibfield  {author} {\bibinfo {author} {\bibfnamefont {X.-L.}\ \bibnamefont
  {Wu}}\ and\ \bibinfo {author} {\bibfnamefont {A.}~\bibnamefont {Libchaber}},\
  }\href@noop {} {\bibfield  {journal} {\bibinfo  {journal} {Physical review
  letters}\ }\textbf {\bibinfo {volume} {84}},\ \bibinfo {pages} {3017}
  (\bibinfo {year} {2000})}\BibitemShut {NoStop}%
\bibitem [{\citenamefont {Elgeti}\ \emph {et~al.}(2015)\citenamefont {Elgeti},
  \citenamefont {Winkler},\ and\ \citenamefont {Gompper}}]{elgeti2015physics}%
  \BibitemOpen
  \bibfield  {author} {\bibinfo {author} {\bibfnamefont {J.}~\bibnamefont
  {Elgeti}}, \bibinfo {author} {\bibfnamefont {R.~G.}\ \bibnamefont {Winkler}},
  \ and\ \bibinfo {author} {\bibfnamefont {G.}~\bibnamefont {Gompper}},\
  }\href@noop {} {\bibfield  {journal} {\bibinfo  {journal} {Reports on
  progress in physics}\ }\textbf {\bibinfo {volume} {78}},\ \bibinfo {pages}
  {056601} (\bibinfo {year} {2015})}\BibitemShut {NoStop}%
\bibitem [{\citenamefont {Prost}\ \emph {et~al.}(2015)\citenamefont {Prost},
  \citenamefont {J{\"u}licher},\ and\ \citenamefont
  {Joanny}}]{prost2015active}%
  \BibitemOpen
  \bibfield  {author} {\bibinfo {author} {\bibfnamefont {J.}~\bibnamefont
  {Prost}}, \bibinfo {author} {\bibfnamefont {F.}~\bibnamefont {J{\"u}licher}},
  \ and\ \bibinfo {author} {\bibfnamefont {J.-F.}\ \bibnamefont {Joanny}},\
  }\href@noop {} {\bibfield  {journal} {\bibinfo  {journal} {Nature physics}\
  }\textbf {\bibinfo {volume} {11}},\ \bibinfo {pages} {111} (\bibinfo {year}
  {2015})}\BibitemShut {NoStop}%
\bibitem [{\citenamefont {Cavagna}\ and\ \citenamefont
  {Giardina}(2014)}]{cavagna2014bird}%
  \BibitemOpen
  \bibfield  {author} {\bibinfo {author} {\bibfnamefont {A.}~\bibnamefont
  {Cavagna}}\ and\ \bibinfo {author} {\bibfnamefont {I.}~\bibnamefont
  {Giardina}},\ }\href@noop {} {\bibfield  {journal} {\bibinfo  {journal}
  {Annu. Rev. Condens. Matter Phys.}\ }\textbf {\bibinfo {volume} {5}},\
  \bibinfo {pages} {183} (\bibinfo {year} {2014})}\BibitemShut {NoStop}%
\bibitem [{\citenamefont {Calovi}\ \emph {et~al.}(2014)\citenamefont {Calovi},
  \citenamefont {Lopez}, \citenamefont {Ngo}, \citenamefont {Sire},
  \citenamefont {Chat{\'e}},\ and\ \citenamefont
  {Theraulaz}}]{calovi2014swarming}%
  \BibitemOpen
  \bibfield  {author} {\bibinfo {author} {\bibfnamefont {D.~S.}\ \bibnamefont
  {Calovi}}, \bibinfo {author} {\bibfnamefont {U.}~\bibnamefont {Lopez}},
  \bibinfo {author} {\bibfnamefont {S.}~\bibnamefont {Ngo}}, \bibinfo {author}
  {\bibfnamefont {C.}~\bibnamefont {Sire}}, \bibinfo {author} {\bibfnamefont
  {H.}~\bibnamefont {Chat{\'e}}}, \ and\ \bibinfo {author} {\bibfnamefont
  {G.}~\bibnamefont {Theraulaz}},\ }\href@noop {} {\bibfield  {journal}
  {\bibinfo  {journal} {New journal of Physics}\ }\textbf {\bibinfo {volume}
  {16}},\ \bibinfo {pages} {015026} (\bibinfo {year} {2014})}\BibitemShut
  {NoStop}%
\bibitem [{\citenamefont {Bottinelli}\ \emph {et~al.}(2016)\citenamefont
  {Bottinelli}, \citenamefont {Sumpter},\ and\ \citenamefont
  {Silverberg}}]{bottinelli2016emergent}%
  \BibitemOpen
  \bibfield  {author} {\bibinfo {author} {\bibfnamefont {A.}~\bibnamefont
  {Bottinelli}}, \bibinfo {author} {\bibfnamefont {D.~T.}\ \bibnamefont
  {Sumpter}}, \ and\ \bibinfo {author} {\bibfnamefont {J.~L.}\ \bibnamefont
  {Silverberg}},\ }\href@noop {} {\bibfield  {journal} {\bibinfo  {journal}
  {Physical review letters}\ }\textbf {\bibinfo {volume} {117}},\ \bibinfo
  {pages} {228301} (\bibinfo {year} {2016})}\BibitemShut {NoStop}%
\bibitem [{\citenamefont {Bain}\ and\ \citenamefont
  {Bartolo}(2019)}]{bain2019dynamic}%
  \BibitemOpen
  \bibfield  {author} {\bibinfo {author} {\bibfnamefont {N.}~\bibnamefont
  {Bain}}\ and\ \bibinfo {author} {\bibfnamefont {D.}~\bibnamefont {Bartolo}},\
  }\href@noop {} {\bibfield  {journal} {\bibinfo  {journal} {Science}\ }\textbf
  {\bibinfo {volume} {363}},\ \bibinfo {pages} {46} (\bibinfo {year}
  {2019})}\BibitemShut {NoStop}%
\bibitem [{\citenamefont {Shaebani}\ \emph {et~al.}(2020)\citenamefont
  {Shaebani}, \citenamefont {Wysocki}, \citenamefont {Winkler}, \citenamefont
  {Gompper},\ and\ \citenamefont {Rieger}}]{shaebani2020computational}%
  \BibitemOpen
  \bibfield  {author} {\bibinfo {author} {\bibfnamefont {M.~R.}\ \bibnamefont
  {Shaebani}}, \bibinfo {author} {\bibfnamefont {A.}~\bibnamefont {Wysocki}},
  \bibinfo {author} {\bibfnamefont {R.~G.}\ \bibnamefont {Winkler}}, \bibinfo
  {author} {\bibfnamefont {G.}~\bibnamefont {Gompper}}, \ and\ \bibinfo
  {author} {\bibfnamefont {H.}~\bibnamefont {Rieger}},\ }\href@noop {}
  {\bibfield  {journal} {\bibinfo  {journal} {Nature Reviews Physics}\ }\textbf
  {\bibinfo {volume} {2}},\ \bibinfo {pages} {181} (\bibinfo {year}
  {2020})}\BibitemShut {NoStop}%
\bibitem [{\citenamefont {Vicsek}\ \emph {et~al.}(1995)\citenamefont {Vicsek},
  \citenamefont {Czir{\'o}k}, \citenamefont {Ben-Jacob}, \citenamefont
  {Cohen},\ and\ \citenamefont {Shochet}}]{vicsek1995novel}%
  \BibitemOpen
  \bibfield  {author} {\bibinfo {author} {\bibfnamefont {T.}~\bibnamefont
  {Vicsek}}, \bibinfo {author} {\bibfnamefont {A.}~\bibnamefont {Czir{\'o}k}},
  \bibinfo {author} {\bibfnamefont {E.}~\bibnamefont {Ben-Jacob}}, \bibinfo
  {author} {\bibfnamefont {I.}~\bibnamefont {Cohen}}, \ and\ \bibinfo {author}
  {\bibfnamefont {O.}~\bibnamefont {Shochet}},\ }\href@noop {} {\bibfield
  {journal} {\bibinfo  {journal} {Physical review letters}\ }\textbf {\bibinfo
  {volume} {75}},\ \bibinfo {pages} {1226} (\bibinfo {year}
  {1995})}\BibitemShut {NoStop}%
\bibitem [{\citenamefont {Solon}\ \emph
  {et~al.}(2015{\natexlab{a}})\citenamefont {Solon}, \citenamefont
  {Chat{\'e}},\ and\ \citenamefont {Tailleur}}]{solon2015phase}%
  \BibitemOpen
  \bibfield  {author} {\bibinfo {author} {\bibfnamefont {A.~P.}\ \bibnamefont
  {Solon}}, \bibinfo {author} {\bibfnamefont {H.}~\bibnamefont {Chat{\'e}}}, \
  and\ \bibinfo {author} {\bibfnamefont {J.}~\bibnamefont {Tailleur}},\
  }\href@noop {} {\bibfield  {journal} {\bibinfo  {journal} {Physical review
  letters}\ }\textbf {\bibinfo {volume} {114}},\ \bibinfo {pages} {068101}
  (\bibinfo {year} {2015}{\natexlab{a}})}\BibitemShut {NoStop}%
\bibitem [{\citenamefont {Solon}\ \emph
  {et~al.}(2015{\natexlab{b}})\citenamefont {Solon}, \citenamefont {Caussin},
  \citenamefont {Bartolo}, \citenamefont {Chat{\'e}},\ and\ \citenamefont
  {Tailleur}}]{solon2015pattern}%
  \BibitemOpen
  \bibfield  {author} {\bibinfo {author} {\bibfnamefont {A.~P.}\ \bibnamefont
  {Solon}}, \bibinfo {author} {\bibfnamefont {J.-B.}\ \bibnamefont {Caussin}},
  \bibinfo {author} {\bibfnamefont {D.}~\bibnamefont {Bartolo}}, \bibinfo
  {author} {\bibfnamefont {H.}~\bibnamefont {Chat{\'e}}}, \ and\ \bibinfo
  {author} {\bibfnamefont {J.}~\bibnamefont {Tailleur}},\ }\href@noop {}
  {\bibfield  {journal} {\bibinfo  {journal} {Physical Review E}\ }\textbf
  {\bibinfo {volume} {92}},\ \bibinfo {pages} {062111} (\bibinfo {year}
  {2015}{\natexlab{b}})}\BibitemShut {NoStop}%
\bibitem [{\citenamefont {Chat{\'e}}\ \emph {et~al.}(2008)\citenamefont
  {Chat{\'e}}, \citenamefont {Ginelli}, \citenamefont {Gr{\'e}goire},\ and\
  \citenamefont {Raynaud}}]{chate2008collective}%
  \BibitemOpen
  \bibfield  {author} {\bibinfo {author} {\bibfnamefont {H.}~\bibnamefont
  {Chat{\'e}}}, \bibinfo {author} {\bibfnamefont {F.}~\bibnamefont {Ginelli}},
  \bibinfo {author} {\bibfnamefont {G.}~\bibnamefont {Gr{\'e}goire}}, \ and\
  \bibinfo {author} {\bibfnamefont {F.}~\bibnamefont {Raynaud}},\ }\href@noop
  {} {\bibfield  {journal} {\bibinfo  {journal} {Physical Review E}\ }\textbf
  {\bibinfo {volume} {77}},\ \bibinfo {pages} {046113} (\bibinfo {year}
  {2008})}\BibitemShut {NoStop}%
\bibitem [{\citenamefont {Xue}\ \emph {et~al.}(2020)\citenamefont {Xue},
  \citenamefont {Li}, \citenamefont {Grassberger},\ and\ \citenamefont
  {Chen}}]{xue2020swarming}%
  \BibitemOpen
  \bibfield  {author} {\bibinfo {author} {\bibfnamefont {T.}~\bibnamefont
  {Xue}}, \bibinfo {author} {\bibfnamefont {X.}~\bibnamefont {Li}}, \bibinfo
  {author} {\bibfnamefont {P.}~\bibnamefont {Grassberger}}, \ and\ \bibinfo
  {author} {\bibfnamefont {L.}~\bibnamefont {Chen}},\ }\href@noop {} {\bibfield
   {journal} {\bibinfo  {journal} {Physical review research}\ }\textbf
  {\bibinfo {volume} {2}},\ \bibinfo {pages} {042017} (\bibinfo {year}
  {2020})}\BibitemShut {NoStop}%
\bibitem [{\citenamefont {Li}\ \emph {et~al.}(2021{\natexlab{a}})\citenamefont
  {Li}, \citenamefont {Xue}, \citenamefont {Sun}, \citenamefont {Fan},
  \citenamefont {Li}, \citenamefont {Liu}, \citenamefont {Han}, \citenamefont
  {Di},\ and\ \citenamefont {Chen}}]{li2021discontinuous}%
  \BibitemOpen
  \bibfield  {author} {\bibinfo {author} {\bibfnamefont {X.}~\bibnamefont
  {Li}}, \bibinfo {author} {\bibfnamefont {T.}~\bibnamefont {Xue}}, \bibinfo
  {author} {\bibfnamefont {Y.}~\bibnamefont {Sun}}, \bibinfo {author}
  {\bibfnamefont {J.}~\bibnamefont {Fan}}, \bibinfo {author} {\bibfnamefont
  {H.}~\bibnamefont {Li}}, \bibinfo {author} {\bibfnamefont {M.}~\bibnamefont
  {Liu}}, \bibinfo {author} {\bibfnamefont {Z.}~\bibnamefont {Han}}, \bibinfo
  {author} {\bibfnamefont {Z.}~\bibnamefont {Di}}, \ and\ \bibinfo {author}
  {\bibfnamefont {X.}~\bibnamefont {Chen}},\ }\href@noop {} {\bibfield
  {journal} {\bibinfo  {journal} {Chinese Physics B}\ }\textbf {\bibinfo
  {volume} {30}},\ \bibinfo {pages} {128703} (\bibinfo {year}
  {2021}{\natexlab{a}})}\BibitemShut {NoStop}%
\bibitem [{\citenamefont {K{\"u}rsten}\ and\ \citenamefont
  {Ihle}(2020)}]{kursten2020dry}%
  \BibitemOpen
  \bibfield  {author} {\bibinfo {author} {\bibfnamefont {R.}~\bibnamefont
  {K{\"u}rsten}}\ and\ \bibinfo {author} {\bibfnamefont {T.}~\bibnamefont
  {Ihle}},\ }\href@noop {} {\bibfield  {journal} {\bibinfo  {journal} {Physical
  Review Letters}\ }\textbf {\bibinfo {volume} {125}},\ \bibinfo {pages}
  {188003} (\bibinfo {year} {2020})}\BibitemShut {NoStop}%
\bibitem [{\citenamefont {K{\"u}rsten}\ \emph {et~al.}(2020)\citenamefont
  {K{\"u}rsten}, \citenamefont {Stroteich}, \citenamefont {H{\'e}rnandez},\
  and\ \citenamefont {Ihle}}]{kursten2020multiple}%
  \BibitemOpen
  \bibfield  {author} {\bibinfo {author} {\bibfnamefont {R.}~\bibnamefont
  {K{\"u}rsten}}, \bibinfo {author} {\bibfnamefont {S.}~\bibnamefont
  {Stroteich}}, \bibinfo {author} {\bibfnamefont {M.~Z.}\ \bibnamefont
  {H{\'e}rnandez}}, \ and\ \bibinfo {author} {\bibfnamefont {T.}~\bibnamefont
  {Ihle}},\ }\href@noop {} {\bibfield  {journal} {\bibinfo  {journal} {Physical
  Review Letters}\ }\textbf {\bibinfo {volume} {124}},\ \bibinfo {pages}
  {088002} (\bibinfo {year} {2020})}\BibitemShut {NoStop}%
\bibitem [{\citenamefont {von Lilienfeld}\ and\ \citenamefont
  {Burke}(2020)}]{von2020retrospective}%
  \BibitemOpen
  \bibfield  {author} {\bibinfo {author} {\bibfnamefont {O.~A.}\ \bibnamefont
  {von Lilienfeld}}\ and\ \bibinfo {author} {\bibfnamefont {K.}~\bibnamefont
  {Burke}},\ }\href@noop {} {\bibfield  {journal} {\bibinfo  {journal} {Nature
  communications}\ }\textbf {\bibinfo {volume} {11}},\ \bibinfo {pages} {1}
  (\bibinfo {year} {2020})}\BibitemShut {NoStop}%
\bibitem [{\citenamefont {Jones}(2017)}]{jones2017machine}%
  \BibitemOpen
  \bibfield  {author} {\bibinfo {author} {\bibfnamefont {N.}~\bibnamefont
  {Jones}},\ }\href@noop {} {\bibfield  {journal} {\bibinfo  {journal}
  {Nature}\ }\textbf {\bibinfo {volume} {548}} (\bibinfo {year}
  {2017})}\BibitemShut {NoStop}%
\bibitem [{\citenamefont {Carleo}\ \emph {et~al.}(2019)\citenamefont {Carleo},
  \citenamefont {Cirac}, \citenamefont {Cranmer}, \citenamefont {Daudet},
  \citenamefont {Schuld}, \citenamefont {Tishby}, \citenamefont
  {Vogt-Maranto},\ and\ \citenamefont {Zdeborov{\'a}}}]{carleo2019machine}%
  \BibitemOpen
  \bibfield  {author} {\bibinfo {author} {\bibfnamefont {G.}~\bibnamefont
  {Carleo}}, \bibinfo {author} {\bibfnamefont {I.}~\bibnamefont {Cirac}},
  \bibinfo {author} {\bibfnamefont {K.}~\bibnamefont {Cranmer}}, \bibinfo
  {author} {\bibfnamefont {L.}~\bibnamefont {Daudet}}, \bibinfo {author}
  {\bibfnamefont {M.}~\bibnamefont {Schuld}}, \bibinfo {author} {\bibfnamefont
  {N.}~\bibnamefont {Tishby}}, \bibinfo {author} {\bibfnamefont
  {L.}~\bibnamefont {Vogt-Maranto}}, \ and\ \bibinfo {author} {\bibfnamefont
  {L.}~\bibnamefont {Zdeborov{\'a}}},\ }\href@noop {} {\bibfield  {journal}
  {\bibinfo  {journal} {Reviews of Modern Physics}\ }\textbf {\bibinfo {volume}
  {91}},\ \bibinfo {pages} {045002} (\bibinfo {year} {2019})}\BibitemShut
  {NoStop}%
\bibitem [{\citenamefont {Jordan}\ and\ \citenamefont
  {Mitchell}(2015)}]{jordan2015machine}%
  \BibitemOpen
  \bibfield  {author} {\bibinfo {author} {\bibfnamefont {M.~I.}\ \bibnamefont
  {Jordan}}\ and\ \bibinfo {author} {\bibfnamefont {T.~M.}\ \bibnamefont
  {Mitchell}},\ }\href@noop {} {\bibfield  {journal} {\bibinfo  {journal}
  {Science}\ }\textbf {\bibinfo {volume} {349}},\ \bibinfo {pages} {255}
  (\bibinfo {year} {2015})}\BibitemShut {NoStop}%
\bibitem [{\citenamefont {Van~Nieuwenburg}\ \emph {et~al.}(2017)\citenamefont
  {Van~Nieuwenburg}, \citenamefont {Liu},\ and\ \citenamefont
  {Huber}}]{van2017learning}%
  \BibitemOpen
  \bibfield  {author} {\bibinfo {author} {\bibfnamefont {E.~P.}\ \bibnamefont
  {Van~Nieuwenburg}}, \bibinfo {author} {\bibfnamefont {Y.-H.}\ \bibnamefont
  {Liu}}, \ and\ \bibinfo {author} {\bibfnamefont {S.~D.}\ \bibnamefont
  {Huber}},\ }\href@noop {} {\bibfield  {journal} {\bibinfo  {journal} {Nature
  Physics}\ }\textbf {\bibinfo {volume} {13}},\ \bibinfo {pages} {435}
  (\bibinfo {year} {2017})}\BibitemShut {NoStop}%
\bibitem [{\citenamefont {Venderley}\ \emph {et~al.}(2018)\citenamefont
  {Venderley}, \citenamefont {Khemani},\ and\ \citenamefont
  {Kim}}]{venderley2018machine}%
  \BibitemOpen
  \bibfield  {author} {\bibinfo {author} {\bibfnamefont {J.}~\bibnamefont
  {Venderley}}, \bibinfo {author} {\bibfnamefont {V.}~\bibnamefont {Khemani}},
  \ and\ \bibinfo {author} {\bibfnamefont {E.-A.}\ \bibnamefont {Kim}},\
  }\href@noop {} {\bibfield  {journal} {\bibinfo  {journal} {Physical review
  letters}\ }\textbf {\bibinfo {volume} {120}},\ \bibinfo {pages} {257204}
  (\bibinfo {year} {2018})}\BibitemShut {NoStop}%
\bibitem [{\citenamefont {Carrasquilla}\ and\ \citenamefont
  {Melko}(2017)}]{carrasquilla2017machine}%
  \BibitemOpen
  \bibfield  {author} {\bibinfo {author} {\bibfnamefont {J.}~\bibnamefont
  {Carrasquilla}}\ and\ \bibinfo {author} {\bibfnamefont {R.~G.}\ \bibnamefont
  {Melko}},\ }\href@noop {} {\bibfield  {journal} {\bibinfo  {journal} {Nature
  Physics}\ }\textbf {\bibinfo {volume} {13}},\ \bibinfo {pages} {431}
  (\bibinfo {year} {2017})}\BibitemShut {NoStop}%
\bibitem [{\citenamefont {Deng}\ \emph {et~al.}(2017)\citenamefont {Deng},
  \citenamefont {Li},\ and\ \citenamefont {Sarma}}]{deng2017quantum}%
  \BibitemOpen
  \bibfield  {author} {\bibinfo {author} {\bibfnamefont {D.-L.}\ \bibnamefont
  {Deng}}, \bibinfo {author} {\bibfnamefont {X.}~\bibnamefont {Li}}, \ and\
  \bibinfo {author} {\bibfnamefont {S.~D.}\ \bibnamefont {Sarma}},\ }\href@noop
  {} {\bibfield  {journal} {\bibinfo  {journal} {Physical Review X}\ }\textbf
  {\bibinfo {volume} {7}},\ \bibinfo {pages} {021021} (\bibinfo {year}
  {2017})}\BibitemShut {NoStop}%
\bibitem [{\citenamefont {Rodriguez-Nieva}\ and\ \citenamefont
  {Scheurer}(2019)}]{rodriguez2019identifying}%
  \BibitemOpen
  \bibfield  {author} {\bibinfo {author} {\bibfnamefont {J.~F.}\ \bibnamefont
  {Rodriguez-Nieva}}\ and\ \bibinfo {author} {\bibfnamefont {M.~S.}\
  \bibnamefont {Scheurer}},\ }\href@noop {} {\bibfield  {journal} {\bibinfo
  {journal} {Nature Physics}\ }\textbf {\bibinfo {volume} {15}},\ \bibinfo
  {pages} {790} (\bibinfo {year} {2019})}\BibitemShut {NoStop}%
\bibitem [{\citenamefont {Wetzel}(2017)}]{wetzel2017unsupervised}%
  \BibitemOpen
  \bibfield  {author} {\bibinfo {author} {\bibfnamefont {S.~J.}\ \bibnamefont
  {Wetzel}},\ }\href@noop {} {\bibfield  {journal} {\bibinfo  {journal}
  {Physical Review E}\ }\textbf {\bibinfo {volume} {96}},\ \bibinfo {pages}
  {022140} (\bibinfo {year} {2017})}\BibitemShut {NoStop}%
\bibitem [{\citenamefont {Zhang}\ \emph {et~al.}(2019)\citenamefont {Zhang},
  \citenamefont {Liu},\ and\ \citenamefont {Wei}}]{zhang2019machine}%
  \BibitemOpen
  \bibfield  {author} {\bibinfo {author} {\bibfnamefont {W.}~\bibnamefont
  {Zhang}}, \bibinfo {author} {\bibfnamefont {J.}~\bibnamefont {Liu}}, \ and\
  \bibinfo {author} {\bibfnamefont {T.-C.}\ \bibnamefont {Wei}},\ }\href@noop
  {} {\bibfield  {journal} {\bibinfo  {journal} {Physical Review E}\ }\textbf
  {\bibinfo {volume} {99}},\ \bibinfo {pages} {032142} (\bibinfo {year}
  {2019})}\BibitemShut {NoStop}%
\bibitem [{\citenamefont {Torlai}\ and\ \citenamefont
  {Melko}(2018)}]{torlai2018latent}%
  \BibitemOpen
  \bibfield  {author} {\bibinfo {author} {\bibfnamefont {G.}~\bibnamefont
  {Torlai}}\ and\ \bibinfo {author} {\bibfnamefont {R.~G.}\ \bibnamefont
  {Melko}},\ }\href@noop {} {\bibfield  {journal} {\bibinfo  {journal}
  {Physical review letters}\ }\textbf {\bibinfo {volume} {120}},\ \bibinfo
  {pages} {240503} (\bibinfo {year} {2018})}\BibitemShut {NoStop}%
\bibitem [{\citenamefont {Ch’Ng}\ \emph {et~al.}(2017)\citenamefont
  {Ch’Ng}, \citenamefont {Carrasquilla}, \citenamefont {Melko},\ and\
  \citenamefont {Khatami}}]{ch2017machine}%
  \BibitemOpen
  \bibfield  {author} {\bibinfo {author} {\bibfnamefont {K.}~\bibnamefont
  {Ch’Ng}}, \bibinfo {author} {\bibfnamefont {J.}~\bibnamefont
  {Carrasquilla}}, \bibinfo {author} {\bibfnamefont {R.~G.}\ \bibnamefont
  {Melko}}, \ and\ \bibinfo {author} {\bibfnamefont {E.}~\bibnamefont
  {Khatami}},\ }\href@noop {} {\bibfield  {journal} {\bibinfo  {journal}
  {Physical Review X}\ }\textbf {\bibinfo {volume} {7}},\ \bibinfo {pages}
  {031038} (\bibinfo {year} {2017})}\BibitemShut {NoStop}%
\bibitem [{\citenamefont {Rem}\ \emph {et~al.}(2019)\citenamefont {Rem},
  \citenamefont {K{\"a}ming}, \citenamefont {Tarnowski}, \citenamefont
  {Asteria}, \citenamefont {Fl{\"a}schner}, \citenamefont {Becker},
  \citenamefont {Sengstock},\ and\ \citenamefont
  {Weitenberg}}]{rem2019identifying}%
  \BibitemOpen
  \bibfield  {author} {\bibinfo {author} {\bibfnamefont {B.~S.}\ \bibnamefont
  {Rem}}, \bibinfo {author} {\bibfnamefont {N.}~\bibnamefont {K{\"a}ming}},
  \bibinfo {author} {\bibfnamefont {M.}~\bibnamefont {Tarnowski}}, \bibinfo
  {author} {\bibfnamefont {L.}~\bibnamefont {Asteria}}, \bibinfo {author}
  {\bibfnamefont {N.}~\bibnamefont {Fl{\"a}schner}}, \bibinfo {author}
  {\bibfnamefont {C.}~\bibnamefont {Becker}}, \bibinfo {author} {\bibfnamefont
  {K.}~\bibnamefont {Sengstock}}, \ and\ \bibinfo {author} {\bibfnamefont
  {C.}~\bibnamefont {Weitenberg}},\ }\href@noop {} {\bibfield  {journal}
  {\bibinfo  {journal} {Nature Physics}\ }\textbf {\bibinfo {volume} {15}},\
  \bibinfo {pages} {917} (\bibinfo {year} {2019})}\BibitemShut {NoStop}%
\bibitem [{\citenamefont {Carleo}\ and\ \citenamefont
  {Troyer}(2017)}]{carleo2017solving}%
  \BibitemOpen
  \bibfield  {author} {\bibinfo {author} {\bibfnamefont {G.}~\bibnamefont
  {Carleo}}\ and\ \bibinfo {author} {\bibfnamefont {M.}~\bibnamefont
  {Troyer}},\ }\href@noop {} {\bibfield  {journal} {\bibinfo  {journal}
  {Science}\ }\textbf {\bibinfo {volume} {355}},\ \bibinfo {pages} {602}
  (\bibinfo {year} {2017})}\BibitemShut {NoStop}%
\bibitem [{\citenamefont {Krizhevsky}\ \emph {et~al.}(2012)\citenamefont
  {Krizhevsky}, \citenamefont {Sutskever},\ and\ \citenamefont
  {Hinton}}]{krizhevsky2012imagenet}%
  \BibitemOpen
  \bibfield  {author} {\bibinfo {author} {\bibfnamefont {A.}~\bibnamefont
  {Krizhevsky}}, \bibinfo {author} {\bibfnamefont {I.}~\bibnamefont
  {Sutskever}}, \ and\ \bibinfo {author} {\bibfnamefont {G.~E.}\ \bibnamefont
  {Hinton}},\ }\href@noop {} {\bibfield  {journal} {\bibinfo  {journal}
  {Advances in neural information processing systems}\ }\textbf {\bibinfo
  {volume} {25}} (\bibinfo {year} {2012})}\BibitemShut {NoStop}%
\bibitem [{SM()}]{SM}%
  \BibitemOpen
  \href@noop {} {\ }\bibinfo {note} {See Supplemental Material for additional
  analytical calculations and numerical results.}\BibitemShut {Stop}%
\bibitem [{\citenamefont {Abadi}\ \emph {et~al.}(2015)\citenamefont {Abadi},
  \citenamefont {Agarwal}, \citenamefont {Barham}, \citenamefont {Brevdo},
  \citenamefont {Chen}, \citenamefont {Citro}, \citenamefont {Corrado},
  \citenamefont {Davis}, \citenamefont {Dean}, \citenamefont {Devin} \emph
  {et~al.}}]{abadi2015tensorflow}%
  \BibitemOpen
  \bibfield  {author} {\bibinfo {author} {\bibfnamefont {M.}~\bibnamefont
  {Abadi}}, \bibinfo {author} {\bibfnamefont {A.}~\bibnamefont {Agarwal}},
  \bibinfo {author} {\bibfnamefont {P.}~\bibnamefont {Barham}}, \bibinfo
  {author} {\bibfnamefont {E.}~\bibnamefont {Brevdo}}, \bibinfo {author}
  {\bibfnamefont {Z.}~\bibnamefont {Chen}}, \bibinfo {author} {\bibfnamefont
  {C.}~\bibnamefont {Citro}}, \bibinfo {author} {\bibfnamefont {G.~S.}\
  \bibnamefont {Corrado}}, \bibinfo {author} {\bibfnamefont {A.}~\bibnamefont
  {Davis}}, \bibinfo {author} {\bibfnamefont {J.}~\bibnamefont {Dean}},
  \bibinfo {author} {\bibfnamefont {M.}~\bibnamefont {Devin}},  \emph
  {et~al.},\ }\href@noop {} {\enquote {\bibinfo {title} {Tensorflow:
  Large-scale machine learning on heterogeneous systems},}\ } (\bibinfo {year}
  {2015})\BibitemShut {NoStop}%
\bibitem [{\citenamefont {LeCun}\ \emph {et~al.}(2015)\citenamefont {LeCun},
  \citenamefont {Bengio},\ and\ \citenamefont {Hinton}}]{lecun2015deep}%
  \BibitemOpen
  \bibfield  {author} {\bibinfo {author} {\bibfnamefont {Y.}~\bibnamefont
  {LeCun}}, \bibinfo {author} {\bibfnamefont {Y.}~\bibnamefont {Bengio}}, \
  and\ \bibinfo {author} {\bibfnamefont {G.}~\bibnamefont {Hinton}},\
  }\href@noop {} {\bibfield  {journal} {\bibinfo  {journal} {nature}\ }\textbf
  {\bibinfo {volume} {521}},\ \bibinfo {pages} {436} (\bibinfo {year}
  {2015})}\BibitemShut {NoStop}%
\bibitem [{\citenamefont {Goodfellow}\ \emph {et~al.}(2016)\citenamefont
  {Goodfellow}, \citenamefont {Bengio},\ and\ \citenamefont
  {Courville}}]{goodfellow2016deep}%
  \BibitemOpen
  \bibfield  {author} {\bibinfo {author} {\bibfnamefont {I.}~\bibnamefont
  {Goodfellow}}, \bibinfo {author} {\bibfnamefont {Y.}~\bibnamefont {Bengio}},
  \ and\ \bibinfo {author} {\bibfnamefont {A.}~\bibnamefont {Courville}},\
  }\href@noop {} {\emph {\bibinfo {title} {Deep learning}}}\ (\bibinfo
  {publisher} {MIT press},\ \bibinfo {year} {2016})\BibitemShut {NoStop}%
\bibitem [{\citenamefont {Shalev-Shwartz}\ and\ \citenamefont
  {Zhang}(2013)}]{shalev2013stochastic}%
  \BibitemOpen
  \bibfield  {author} {\bibinfo {author} {\bibfnamefont {S.}~\bibnamefont
  {Shalev-Shwartz}}\ and\ \bibinfo {author} {\bibfnamefont {T.}~\bibnamefont
  {Zhang}},\ }\href@noop {} {\bibfield  {journal} {\bibinfo  {journal} {Journal
  of Machine Learning Research}\ }\textbf {\bibinfo {volume} {14}} (\bibinfo
  {year} {2013})}\BibitemShut {NoStop}%
\bibitem [{\citenamefont {Ni}\ \emph {et~al.}(2019)\citenamefont {Ni},
  \citenamefont {Tang}, \citenamefont {Liu},\ and\ \citenamefont
  {Lai}}]{ni2019machine}%
  \BibitemOpen
  \bibfield  {author} {\bibinfo {author} {\bibfnamefont {Q.}~\bibnamefont
  {Ni}}, \bibinfo {author} {\bibfnamefont {M.}~\bibnamefont {Tang}}, \bibinfo
  {author} {\bibfnamefont {Y.}~\bibnamefont {Liu}}, \ and\ \bibinfo {author}
  {\bibfnamefont {Y.-C.}\ \bibnamefont {Lai}},\ }\href@noop {} {\bibfield
  {journal} {\bibinfo  {journal} {Physical Review E}\ }\textbf {\bibinfo
  {volume} {100}},\ \bibinfo {pages} {052312} (\bibinfo {year}
  {2019})}\BibitemShut {NoStop}%
\bibitem [{\citenamefont {Li}\ \emph {et~al.}(2021{\natexlab{b}})\citenamefont
  {Li}, \citenamefont {Jin}, \citenamefont {Jiang},\ and\ \citenamefont
  {Chen}}]{li2021determining}%
  \BibitemOpen
  \bibfield  {author} {\bibinfo {author} {\bibfnamefont {H.}~\bibnamefont
  {Li}}, \bibinfo {author} {\bibfnamefont {Y.}~\bibnamefont {Jin}}, \bibinfo
  {author} {\bibfnamefont {Y.}~\bibnamefont {Jiang}}, \ and\ \bibinfo {author}
  {\bibfnamefont {J.~Z.}\ \bibnamefont {Chen}},\ }\href@noop {} {\bibfield
  {journal} {\bibinfo  {journal} {Proceedings of the National Academy of
  Sciences}\ }\textbf {\bibinfo {volume} {118}},\ \bibinfo {pages}
  {e2017392118} (\bibinfo {year} {2021}{\natexlab{b}})}\BibitemShut {NoStop}%
\end{thebibliography}%
\end{document}